\documentclass[%
  twoside,
  floatfix,
  reprint,
  amsmath,amssymb,
  aps,
  pra,
  nofootinbib,
  showpacs,
  superscriptaddress,
  a4paper
]{revtex4-1}

\usepackage{graphicx}%
\usepackage[usenames,dvipsnames]{xcolor}
\usepackage{siunitx}
\usepackage{subfigure}
\usepackage{enumitem}
\usepackage{subfigure}
\usepackage{ulem}

\usepackage{tabularx}
\usepackage{booktabs}
\usepackage{multirow}
\usepackage{soul}
\setstcolor{red}
\usepackage{titlesec}

\usepackage{array}

\usepackage[utf8]{inputenc}
\usepackage[T1]{fontenc}
\usepackage{ifpdf}

\usepackage{lipsum}
\graphicspath{{Figures/}{}}

\usepackage[
centering, includefoot,
text={7.1in,10.2in},
total={6.3in,8.75in},
top=0.8in, left=0.62in,
]{geometry}

\usepackage[
  bookmarks=false,
  colorlinks,
  linkcolor=blue,
  urlcolor=blue,
  citecolor=blue,
  plainpages=false,
  pdfpagelabels,
  final,
  breaklinks=true
]{hyperref}
\hypersetup{
pdftitle={Laser-assisted photoionization: beyond the dipole approximation}, 
pdfauthor={R. Della Picca, et al}
}

\usepackage{natbib}
\makeatletter \def\NAT@def@citea{\def\@citea{\NAT@separator\,}} \makeatother

\newcommand{\orcid}[1]{%
  \href{%
    https://orcid.org/#1%
  }{%
   \,\protect\includegraphics[width=8pt]{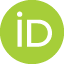}%
  }%
}

\usepackage{physics}
  \newcommand{\vbr}{\vb{r}}
  \newcommand{\vbp}{\vb{p}}
  \newcommand{\vbk}{\vb{k}}
  \newcommand{\vba}{\vb{A}}
  \newcommand{\vbA}{\vb{A}}
  \newcommand{\vbf}{\vb{F}}
  \newcommand{\vbF}{\vb{F}}
  \newcommand{\vbK}{\vb{K}}

  \newcommand{\vbd}{\vb{d}}
  
  \newcommand{\vbv}{\vb{v}}
 \newcommand{\rme}{\mathrm{e}}
\newcommand{\rmd}{\mathrm{d}}

\newcommand{\Eref}[1]{Equation (\ref{#1})}
\newcommand{\eref}[1]{Eq.~(\ref{#1})}

\newcommand{\infi}{\ensuremath{\mathrm{if}}}

\DeclareSIUnit{\au}{{a.u.}}

\newcommand{\nhphantom}[1]{\sbox0{#1}\hspace{-\the\wd0}}

\begin{document}
\title{Laser-assisted photoionization:\\
beyond the dipole approximation}

\author{R. Della Picca\,\orcid{0000-0001-7909-4529}}
\email[]{renata@cab.cnea.gov.ar}
\affiliation{Centro At\'omico Bariloche (CNEA),  CONICET and Instituto Balseiro (UNCuyo), 8400 Bariloche, Argentina}

\author{J. M. Randazzo}
\affiliation{Centro At\'omico Bariloche (CNEA),  CONICET and Instituto Balseiro (UNCuyo), 8400 Bariloche, Argentina}

\author{S. D. L\'opez}
\affiliation{Institute for Astronomy and Space Physics - IAFE (UBA-Conicet), Buenos Aires, Argentina}

\author{M. F. Ciappina\,\orcid{0000-0002-1123-6460}}
\email{marcelo.ciappina@gtiit.edu.cn}
\affiliation{Department of Physics, Guangdong Technion - Israel Institute of Technology, 241 Daxue Road, Shantou, Guangdong, China, 515063}
\affiliation{Technion -- Israel Institute of Technology, Haifa, 32000, Israel}
\affiliation{Guangdong Provincial Key Laboratory of Materials and Technologies for Energy Conversion, Guangdong Technion - Israel Institute of Technology, 241 Daxue Road, Shantou, Guangdong, China, 515063}

\author{D. G. Arb\'o\,\orcid{0000-0002-4375-4940}}
\affiliation{Institute for Astronomy and Space Physics - IAFE (UBA-Conicet), Buenos Aires, Argentina}
\affiliation{Universidad de Buenos Aires - Facultad de Ciencias Exactas y Naturales and Ciclo B\'asico Com\'un , Buenos Aires, Argentina}

\date{\today}

\begin{abstract}
We present a theoretical study of atomic laser-assisted photoionization emission (LAPE) beyond the dipole approximation. By considering the non-relativistic non-dipole strong-field approximation (non-dipole Gordon-Volkov wave function), 
we analyze the different contributions to the photoelectron spectrum (PES), which can be written in terms of intra- and intercycle factors. 
We find that not only does our non-dipole approach exhibit asymmetric emission in the direction of light propagation, but also allows emission in dipole-forbidden directions. 
The former feature can be rooted both in intra- and intercycle interference processes, whilst the latter stems from a dependence of the sideband energy on the emission angle with respect to the propagation direction.
Our theoretical scheme, presented here for He atoms in the $1s$ quantum state, is general enough to be applied to other atomic species and field configurations.
\end{abstract}

\pacs{32.80.Wr, 32.80.Fb, 03.65.Sq}
\maketitle


\section{Introduction}

When an extreme ultraviolet (XUV) pulse and an infrared (IR) laser field overlap in space and time with matter, the so-called laser-assisted photoionization emission (LAPE) processes take place. Here, two main and distinct schemes can be distinguished depending on the XUV pulse duration, namely (i) the streaking regime, when the XUV pulse duration is shorter than one IR optical cycle; and (ii) the sideband regime, when the XUV pulse duration is longer than it. The XUV pulse promotes an electron wavepacket to the continuum in the presence of the IR field. In (i), if both the XUV and IR fields are controlled with sub-femtosecond time resolution, the photoelectron spectra for different time delays can be recorded. This \textit{spectrogram} carries structural, amplitude, and phase information of both the XUV and IR fields. These parameters can be efficiently retrieved by applying well-established reconstruction algorithms~\cite{Mairesse2005,Goulielmakis2008,Goulielmakis2004,Gagnon2009}.
Alternatively, in (ii), the concurrent absorption of one XUV photon, together with the exchange of one or more additional photons from the IR laser field, leads to  equally spaced ``sideband'' (SB) peaks in the energy-resolved photoelectron spectrum (PES). They are located at higher and lower energies than the XUV photoionization energy value due to respective absorption and emission of IR photons~\cite{Drescher2005,Meyer2010}.
The first theoretical prediction of the sideband (SB) peaks was presented in~\cite{Veniard95}. Since then, a great deal of experimental and theoretical work has been reported in this field (see, e.g.,~\cite{Itatani02, Drescher2005, Maquet2007, Radcliffe2012, Meyer2012, Mazza2014,Dusterer2019} and references therein). 

The sideband peaks have great similarities with 
the well-known Above Threshold Ionization (ATI) peaks in the context of multiphoton strong-field ionization, where the target absorbs more photons than those required for an electron to get ionized~\cite{Milosevic2006}.
The formation of both kinds of peaks (ATI and SB) can be theoretically explained as the constructive interference between electron wavepackets released at different optical cycles of the IR laser field~\cite{Arbo2010a, Kazansky10b,Gramajo18}.

Previously, we successfully identified electron trajectories and described the PES as a transparent product of inter- and intracycle interference factors based on the strong-field approximation (SFA) within the dipole approximation for both ATI \cite{Arbo2010a, Arbo2010b, Arbo2012} and LAPE scenarios~\cite{Gramajo18, Gramajo16,Gramajo17}. We showed that the intercycle interference accounts for the sidebands' formation and the intracycle interference appears as a modulation of the former. 
 These two types of interference  can be easily explained as the coherent superposition of electron trajectories making use of the saddle-point approximation (SPA), for the calculation (time integration) of the transition matrix. However, resorting to the SPA is not necessary as we have shown in Refs. \cite{DellaPicca2020, proceedingIcpeac19}, where we have demonstrated 
 that it is feasible to compute the PES as a function of a kernel quantity that represents the  time-dependent photoionization transition matrix for an (only one IR cycle duration) XUV pulse.

In all the above-described theoretical approaches the dipole approximation was considered provided (i) the IR laser is weak enough and (ii) laser wavelengths are short enough that considering magnetic effects can be neglected. 
Within the dipole approximation, photons transmit energy to the target (atom, molecule, or solid) but not momentum since the laser electric field is considered homogeneous with no contribution of the magnetic field component.
These assumptions must be revised when (i) ultrastrong laser fields~\cite{P3ELI} and (ii) mid-IR laser sources~\cite{Elu2019,Elu2021} are used.
For long-wavelength high-intensity lasers, non-dipole effects originate from the Lorentz force of the magnetic field and give rise to a momentum transfer of laser photons on the ejected electrons along the propagation direction. The contribution of relativistic effects in ATI can be quantified through the parameter \cite{Reiss08a,Reiss08b,Reiss14}
\begin{equation}
  q=\frac{U_p}{mc^2},   
\end{equation}
that reflects the importance of the ponderomotive energy $U_p$, relative to the rest energy of the electron ($m$ is the electron rest mass and $c$ the speed of light). For small $q$, and when the Coulombic effect of the remaining core on the photoelectron is disregarded, the classical motion of the ejected electron can be thought of as a composition of two motions:
(i) the electron oscillates in the polarization direction due to the (dipole) laser electric field with the aforementioned ponderomotive energy and (ii) the electron drifts along the light propagation direction superimposed with an oscillation with twice the laser frequency with the well-known ``figure eight'' motion, in a reference frame accompanying the electron in its average drift motion. The drift per cycle relative to the (dipole) quiver amplitude can be quantified as $\pi \sqrt{q}$ and the amplitude $\beta_0$ of the ``figure eight'' motion along the propagation direction relative to the amplitude along the polarization direction as $\sqrt{q}/4$. A momentum shift along the laser propagation direction at the tunnel exit is a signature of the relativistic dynamics through the tunneling barrier in above-threshold ionization (ATI) \cite{Chirila02}. Non-dipole effects break the forward/backward symmetry of electron emission strongly reducing recollision in high-order above-threshold ionization (HATI) affecting, therefore, photon emission by laser-driven ions \cite{Brennecke18}. 

The partitioning of the photon momentum transfer between electron and ion is currently under debate  \cite{Klaiber13,Brennecke18,NI20,Fritzsche22,Maurer21,Kahvedzic22,Popov04,Jiang22,Lin22}. 
Part of the photon momentum shift that the electron takes stems from the effect of the magnetic field during tunneling through the potential barrier formed by the atomic potential and the laser field.
The remaining part of the photon momentum is transferred to the electron during its motion in the continuum \cite{Chelkovski14,Willenberg19,Hartung19,Lin22}. For linear polarization, a shift of the low-energy region of the momentum distribution against the propagation direction has been reported \cite{Brennecke18,Kahvedzic22,Lin22}.
Recently, it was shown that the subcycle linear momentum transfer can be explained through the interplay between nondipole and nonadiabatic effects on the tunneling dynamics \cite{NI20}.
Comparisons with experiments show that the laser beam profile must be considered together with non-dipole effects to accurately describe the energy of the ATI peaks \cite{Hartung19,Boening21}.

Despite the great and recent research activity of non-dipole effects in strong field ionization (see the works cited in the previous paragraph), to our knowledge there is currently no investigation of how the leading-order non-dipole corrections affect the LAPE scenario.
In this context, we consider (i) the absorption of one XUV photon followed by (ii) multiple absorption or emission of IR photons. Therefore, there is no possibility of momentum transfer of the IR photon to the atom at stage (i) and the IR photon momentum is only transferred during the photoelectron excursion in the continuum (ii).
For this reason, in the current contribution, we study non-dipole effects in  LAPE. 
The aim of the present work  is to unravel the non-dipole traces in the PES structures, that are encoded in both, the intra- and intercycle interference patterns. 
We also analyze how non-dipole effects shift the borders of the classically allowed region.

The paper is organized as follows: In Sec. \ref{sec:level2}, we briefly resume the leading-order non-dipole SFA theory and analyze the properties of the temporal integral of the transition matrix.
In Sec. \ref{sec:level2}A we analyze the intercycle contribution, then in Sec. \ref{sec:level2}B we consider the intracycle factor, and, finally, in  Sec. \ref{sec:level2}C we analyze the semiclassical model for LAPE under non-dipole conditions. In all cases, 
we scrutinize on the asymmetry of the forward-backward emissions. 
Concluding remarks are presented in Sec.~\ref{conclusions}.
Atomic units are used throughout the paper, except where otherwise stated.

\section{\label{sec:level2}Theory and results} 

In the single-active-electron (SAE) approximation, the time-dependent Schr\"odinger equation (TDSE) reads
\begin{equation}
i\frac{\partial }{\partial t}\left\vert \psi (t)\right\rangle =
\Big[ H_0  + H_\textrm{int}(t)  \Big]
\left\vert
\psi (t)\right\rangle , 
\label{TDSE}
\end{equation}%
where $H_0=\vbp^2/2+V(r)$ is the time-independent atomic Hamiltonian, whose first term
corresponds to the electron kinetic energy, and its second term to the electron-core Coulomb interaction.
The second term in the right-hand side of \eref{TDSE}, i.e.,  
\begin{equation}
H_\textrm{int}=\big[ \vbr -iz/c \nabla \big]  \cdot \big( \vbf_{X}(\eta) +  \vbf_{L}(\eta)\big),
\end{equation}
describes the interaction of the atom with both time-dependent XUV [$\vbf_{X}(\eta)$] and IR [$\vbf_{L}(\eta)$] electric fields in the length gauge, with $\eta= \eta (t,\vbr)$
\cite{Joachain12}. 
We suppose the XUV pulse to be weak enough and of short wavelength so that XUV ionization
can be regarded within the dipole approximation, leaving the non-dipole effects to the subsequent action of the NIR laser, i.e., $\vbf_{X}(\eta) = \vbf_{X}(\omega_X t)$.

The electron initially bound in an atomic state $|\phi_{i}\rangle$ is emitted to a final continuum state $|\phi_{f}\rangle$, with final momentum $\vbk$ and energy $E=k^2/2$. 
Then, the energy and angle-resolved photoelectron spectra (PES) can be calculated as
\begin{equation}
\frac{\rmd P}{\rmd E \rmd \Omega}=\sqrt{2 E}\ |T_{\infi}|^2 ,
\label{prob1}
\end{equation}
where $T_{\infi}$ is the $T$-matrix element corresponding to the transition $\phi_{i}\rightarrow\phi_{f}$ and $\rmd\Omega=\sin\theta\rmd\theta\rmd\phi$ is the solid angle, with $\theta$ and $\phi$ the polar and azimuthal angles of the laser-ionized electron, respectively.

Within the time-dependent distorted wave theory, the transition amplitude in the prior form is expressed as
\begin{equation}
T_{\infi}= -i\int_{-\infty}^{+\infty}\rmd t \,\langle\chi_{f}^{-}(\vbr,t)|H_\textrm{int}(\vbr,t)|\phi_{i}(\vbr,t)\rangle,
\label{Tif}
\end{equation}
where $\phi_{i}(\vbr,t)=\varphi_{i}(\vbr)\,e^{i I_{p} t}$ is the initial atomic state, with
ionization potential $I_{p}$, and $\chi_{f}^{-}(\vbr,t)$ is the distorted final state.
\Eref{Tif} is exact as far as the final channel, $\chi_{f}^{-}(\vbr,t)$, is the exact solution of \eref{TDSE}.
However, several degrees of approximation have been considered so far to solve \eref{Tif}. The widest-known one is the SFA, which neglects the Coulomb distortion in the final channel
produced on the ejected-electron state due to its interaction with the residual ion and discard the influence of the laser field in the initial ground state~\cite{maciej1994,symphony}. 
The SFA, for instance, is able to model the `ring' structures of the ATI photoelectron spectrum~\cite{maciej1995}.

In this work, we consider the ionization of an atomic system by the combination of an XUV finite laser pulse assisted by an IR laser polarized in the $\hat{x},\hat{y}$ plane ($\hat{\varepsilon}_L$) and propagating in  the $\hat{z} $ direction with wave vector $\vbK_L = K_L \hat{z}$. We describe the space- and time-dependent IR laser pulse by the vector potential as (see Section 2.8 of \cite{Joachain12}): 
\begin{equation}
\vbA_L(\vbr,t)=\vbA_L(\eta) = \hat{\varepsilon}_L    A_L(\eta), \end{equation}
where $\eta = \omega_L t - \vbK_L \cdot \vbr = \omega_L (t-z/c)$ and the corresponding electric field is:
\begin{equation}
    \vbF_L(\eta) = -\frac{\partial}{\partial t} \vbA_L(\eta) = \hat{\varepsilon}_L F_L(\eta).
\end{equation}

We are interested in the non-dipole effects on the LAPE processes, which let us consider a space-dependent laser field at the lowest order in $1/c$ for the vector potential
\begin{eqnarray}
\vbA_L(\eta) 
&\simeq& \vbA_L(\eta)\big|_{\vbr=0} + (\vbr \cdot \nabla) \vbA_L(\eta)\big|_{\vbr =0} \nonumber \\
&\simeq&\vbA_L(\omega_L t) + \frac{z}{c} \vbF_L(\omega_L t),
\end{eqnarray}
where $\vbA_L(\eta)\big|_{\vbr=0}=\vbA_L(\omega_L t)$. 
Then we approximate the distorted final state with the non-dipole Gordon-Volkov wave function in the length gauge (see Eq.~(2.199) of \cite{Joachain12}):
\begin{eqnarray}
 \chi_{f}^{V_{\textrm{ND}}}(\vbr,t) &=& 
 (2\pi)^{-3/2} \exp \{ i\, \boldsymbol{\Pi}(\vbk,t) \cdot \vbr \}\nonumber \\ 
&&\times \exp{ \frac{i}{2} \int_{t}^{\infty} \boldsymbol{\Pi}^2(\vbk,t^\prime) \rmd t^\prime }, \label{volkovNONdipo}
\end{eqnarray}
where
\begin{equation}
\boldsymbol{\Pi}(\vbk,t) = \vbk + \vbA_L(\omega_L t) + \big[ \vbk \cdot \vbA_L(\omega_L t) + \frac{1}{2} \vbA_L^2(\omega_L t) \big] \frac{\hat{z}}{c}, \label{pi}
\end{equation}
and
\begin{eqnarray}
\boldsymbol{\Pi}^2(\vbk,t) &=& k^2 +
2(\vbk\cdot\hat{\varepsilon}_L) [1+ \frac{\vbk \cdot \hat{z}}{c}] A_L(\omega_L t) \nonumber \\
&& + [1 + \frac{\vbk\cdot\hat{z}}{c} + \frac{(\vbk\cdot\hat{\varepsilon}_L)^2}{c^2}] A^2_L(\omega_L t) \nonumber \\
&& + \frac{\vbk\cdot\hat{\varepsilon}_L}{c^2} A_L^3(\omega_L t) 
   +\frac{1}{4c^2}A_L^4(\omega_L t). \label{Pi2}
\end{eqnarray}
Here, we have taken into account that the IR contribution to the vector potential is dominant.
As the frequency of the XUV pulse is much higher than the one of the IR field,
and considering the strength of the XUV field is much smaller than the IR one, the XUV contribution to the vector potential can be neglected~\cite{Nagele11, DellaPicca13}. 
Within the dipole approximation (in the $1/c \rightarrow 0$ limit), we can approximate the distorted final state with a Volkov function, which is the solution of
the TDSE for a free electron in a homogeneous electromagnetic field
\cite{Volkov}.

With the appropriate choice of the IR and XUV laser parameters, we can assume that the energy domain of the LAPE processes is well separated from the domain of ionization by the IR laser alone. 
In other words, the contribution of IR ionization is negligible in the energy domain where the absorption of one XUV photon takes place, then $H_\textrm{int} \simeq \vbr \cdot \vbF_X$.
Besides, we set the general expression for the linearly polarized XUV pulse of duration $\tau_X$
as 
\begin{equation}
\vbf_{X}(\omega_L t)=-\hat{\varepsilon}_X F_{X0}(t)\cos(\omega_{X}t),
\end{equation}
where $\hat{\varepsilon}_X$ and $\omega _{X}$ are the polarization vector and the carrier frequency of the XUV field, respectively. Furthermore, $F_{X0}(t)$ is a nonzero envelope function during the temporal interval $(t_{0},t_{0}+\tau_{X})$ and zero otherwise, which we approximate as its maximum amplitude, i.e.~$F_{X0}(t)\approx F_{X0}$. Thus, the matrix element of \eref{Tif} can be written as
\begin{eqnarray}
T_{\infi} & = & \int_{t_0}^{t_0+\tau_X} \ell(t) 
\, \rme^{iS(t)} \,\,\rmd t, \label{Tifg}
\end{eqnarray}
where $S(t)$ is the generalized action 
\begin{equation}
S(t)= (I_p - \omega_X) t + \frac{1}{2} \int^{t}  \boldsymbol{\Pi}^2(\vbk,t^\prime)  \, \rmd t' ,
\label{S1}
\end{equation}
and
\begin{equation}
\ell(t) =- \frac{i}{2} F_{X0}\hat{\varepsilon}_X \cdot\vbd \big[ \boldsymbol{\Pi}(\vbk,t) \big],
\label{lt}  
\end{equation}
with the  dipole  moment defined as $\vbd(\vbv)=  (2\pi)^{-3/2}\langle e^{i \vbv \cdot \vbr} \vert \vbr \vert \varphi_{i}(\vbr) \rangle$ (see Appendix). In \eref{Tifg} we have used the rotating wave approximation (RWA) which accounts, in this case, for the absorption of only one XUV photon and neglects, thus,  
the contribution of XUV photoemission.
In addition, during the time-lapse the XUV pulse is acting,
the IR linearly polarized electric field can be modeled as a cosine-like wave, hence, the
vector potential can be written as
\begin{equation}
 \vba_L(\omega_L t)=\frac{F_{L0}}{\omega_L}~\sin{(\omega_{L} t)}~\hat{\varepsilon}_L \label{Avector},
\end{equation}
where $F_{L0}$ is the electric field peak amplitude.
Considering the $T$-periodicity of the vector potential in \eref{Avector}, i.e., $T=2\pi/\omega_L$, and the $\boldsymbol{\Pi}$ dependence on time through $\vbA_L(\omega_L t)$, the dipole moment also fulfills, 
\begin{equation}
\vbd \big[ \boldsymbol{\Pi}(t+jT)\big] =\vbd \big[ \boldsymbol{\Pi}(t)\big],
\label{dipN}
\end{equation} 
for each integer number $j$.

Let us now analyze some features of the $T$-matrix, \eref{Tifg}. To this end we notice that the action $S(t)$ defined in \eref{S1} 
can be written as:
\begin{eqnarray}
S(t) &=& a t + b \cos (\omega_L t) + f \sin (2\omega_L t) \nonumber \\
&& + d \cos (3\omega_L t) + e \sin (4 \omega_L t)
\label{S},
\end{eqnarray}
where
\begin{eqnarray}
a &=&   \frac{k^{2}}{2}+I_{p} - \omega_{X} +U_p \Big[1+\frac{\vbk\cdot\hat{z}}{c}+ \frac{(\vbk\cdot\hat{\varepsilon}_L)^2}{c^2} +\frac{3 U_p}{4c^2}  \Big], \nonumber \\
b &=&  \frac{-F_{L0}}{\omega_{L}^{2}} (\vbk \cdot \hat{\varepsilon}_L) \Big[ 1+\frac{\vbk \cdot \hat{z}}{c}+ \frac{3 U_p}{2c^2} \Big] , \nonumber \\
f &=&\frac{- U_p }{2\omega_{L}}  \Big[1+\frac{\vbk\cdot\hat{z}}{c}+ \frac{(\vbk\cdot\hat{\varepsilon}_L)^2}{c^2} +\frac{U_p}{c^2} \Big], \label{abc}  \\
d &=& (\vbk \cdot \hat{\varepsilon}_L) \frac{ F_{L0} U_p}{6 \omega^2 c^2} \nonumber \\
e &=& \frac{U_p ^4}{16 \omega c^2}. \nonumber
\end{eqnarray}
Here, $U_p=(F_{L0} /2 \omega_L)^2$ is the ponderomotive energy for homogeneous fields.

We then observe that $[S(t)-a t]$ is a time-oscillating function with the same period $T$ of the IR laser field, i.e.,
\begin{equation}
S(t+j T) = S(t) + a j T.
\label{S_N}
\end{equation}
In light of these periodicity properties [Eqs.~(\ref{dipN}) and (\ref{S_N})], we can rewrite the transition matrix [\eref{Tifg}] in terms of the contribution of the first IR cycle only, as we have demonstrated in \cite{DellaPicca2020} within the dipole approximation. 
For that, let us introduce the kernel quantity $I(t)$, as the contribution to the transition amplitude from zero to time $t$, i.e.
\begin{equation}
I(t) = \int_{0}^{t}
\ell(t^\prime) \, \rme^{iS(t^\prime)} \, \rmd t^\prime,  \label{It}
\end{equation}
providing that $0\leq t\leq T$.
From its proper definition, 
it is clear that $I(t)$ increases from zero at $t=0$ and depends on both the electron energy and the geometrical arrangement between $ \hat{\varepsilon}_X $,  $ \hat{\varepsilon}_L $, $\hat{z}$ and the electron emission direction $ \hat{k}$. 

In a previous work \cite{DellaPicca2020} we have presented the expression for the transition matrix as a function of the kernel quantity $I$ for several cases of LAPE processes: streaking, sideband, and pulse train regimes. 
Since the development of those formulas is based on the same periodicity properties of 
 Eqs.~(\ref{dipN}) and (\ref{S_N}) in the present work, we finally find that in the non-dipole situation the Eqs.~(18),  (21), (22), (23), (33) and (36) of \cite{DellaPicca2020} remain valid. 
 Among the three possible regimes, the sideband (SB) scenario is the most relevant
since it is described as the product of two kinds of interferences: the intra- and intercycle contributions,  whilst the other cases can be interpreted in light of these two factors but in a more complicated 
formula than a simple product.
For that, 
in the present work, we focus only on the SB regime (setting no XUV delay, i.e. $t_0=0$ and $\tau_X=N T$ with $N$ an integer number), where the PES is proportional to [see Eq.~(23) of \cite{DellaPicca2020}]:
\begin{equation}
|T_{\infi}^{SB}|^2 =  \underbrace{|I(T)|^{2}}_{\textrm{intracycle}} ~
\underbrace{\left[\frac{\sin{( a T N /2)}}{ \sin{(a T/2) }}\right]^2}_{\textrm{intercycle}}  .
\label{intrainter}
\end{equation} 
In the following, we analyze both, the intra- and intercycle contributions beyond the dipole approximation. However,  we keep our approach
under the condition $ q \ll 1 $, i.e. a nonrelativistic description including nondipole effects.
For this reason in \eref{Pi2} and (\ref{abc}), the terms of order $1/c^2$ will be neglected and only terms 
 proportional to $ \hat{z}/c$ are to be incorporated as a correction to the dipole approximation. 

We consider the ionization of a He$(1s)$ as an example and we fix the IR polarization vector  in  $\hat{x}$ and the XUV one parallel to the IR propagation direction $\hat{z}$, i.e.,  
\[ \hat{\varepsilon}_L=\hat{x}, \qquad \hat{\varepsilon}_X=\hat{z}. \]
In table \ref{t1} we show the rest of the laser parameters for several study cases, chosen in such a way that $\omega_X-I_p-U_p = 2.596$, corresponding to the 0th-order SB position,
be identical for all cases.

\begin{table*}
\caption{Laser parameters in atomic units for each study case. $F_{X0}=0.01$ and $\tau_X=NT$ with $N=1$ or 2. }
\label{t1}
\begin{center}
\begin{tabular}{|c|c|c|c|c|c|c|}
\hline
  CASE&$F_{L0}$& $\omega_L $ ($\lambda_L$ in nm) &$U_p$&$\omega_X$& $\sqrt{q}$&$\beta_0=q ~ c/2\omega_L $\\
\hline

\hline
A & $0.05$ & $0.05$ (911.26 nm) &$1/4$   &  $3.75$ & $0.5/c$ & $2.5/c$ \\
B & $0.05/\sqrt{5}\simeq0.02$ & $0.05/\sqrt{5}$ (2037.6 nm)  &$1/4$   &  $3.75$ & $0.5/c$ & $5.6/c$\\
C & $0.05/\sqrt{5}\simeq0.02$ & $0.05/\sqrt{10}$ (2881.6 nm) &$1/2$ & 4.0 & $0.7/c$ & $16/c$\\
D & $0.05/\sqrt{5}\simeq0.02$ & $0.05/\sqrt{20}$ (4075.3 nm) &$1$ & 4.5 & $1/c$ & $45/c$\\
E & $0.05/\sqrt{2}\simeq0.03$ & $0.05/\sqrt{12}$  (3156.7 nm) &$3/2$ & 5 & $1.2/c$ & $52/c$\\
\hline
\end{tabular}
\end{center}
\end{table*}

\subsection{Intercycle factor}
The zeros of the denominator in the intercycle factor, i.e., the energy values satisfying $a T/2 =n\pi$, are avoidable singularities since the numerator also cancels out and maxima are reached at these points. Such maxima are recognized as the sidebands peaks in the PES.
In the $1/c\rightarrow0$ limit, i.e., within the dipole approximation, the sideband peak of order $n$ occurs at 
$ E_n^D = n \omega_L + \omega_X -I_p-U_p, \label{SBdip} $
corresponding to the absorption ($n>0$) or emission ($n<0$) of $n$ IR photons, following the absorption of one XUV photon.
However, in the present case $aT/2=n\pi$ leads to a quadratic equation for $k_n=\sqrt{2 E_n}$:
\begin{eqnarray}
E_n &= & n \omega_L + \omega_X -I_p-U_p \Big[ 1 + \frac{\vbk_n\cdot \hat{z}}{c} \Big] . \label{sidebandND}
\end{eqnarray}
Rewriting this equation in terms of the parallel ($k_{nz}$)  and perpendicular ($k_{n\perp}$) components of the momentum electron with respect to the propagation direction, we find
\begin{equation}
   \frac{k_{n\perp}^2}{2} + \frac{1}{2}\Big(k_{nz} +\frac{U_p}{c}\Big)^2 \simeq n \omega_L + \omega_X -I_p-U_p , 
\end{equation}
to order $O(1/c)$, which can be understood as a ring in the momentum space with radius $\sqrt{2(n\omega_L +\omega_X-I_p-U_p)}$ and shifted an amount $-U_p/c$ in the $\hat{z}$ direction.
Let us note that a similar result is obtained for the ATI peaks, see for example \cite{Kahvedzic22}. 
We observe that the new positions of the SB peaks are dependent on the projection of the emission direction into the IR propagation axis:  $\vbk\cdot\hat{z} = k \cos\theta $.

\begin{figure}[btp]
\centering
\includegraphics[angle = 0, width=0.45 
\textwidth]{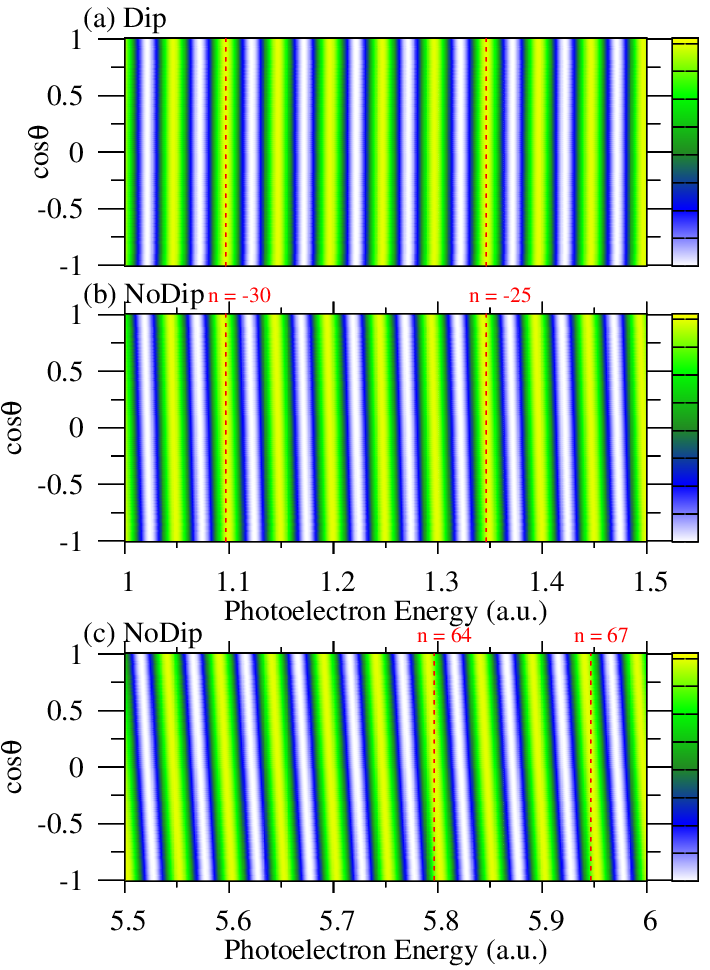}
\caption{(Color online) Intercycle factor for case A (see Table~\ref{t1}) parameters and  $N=2$, within (a) dipole, (b) and (c) non-dipole approximation.
The vertical dashed lines indicate the position of the SB in the dipole approximation: $E_n^D$ for $n= -30, -25$ in (a) and (b); and $n=64, 67$ in (c).
}
\label{sideband}
\end{figure}

In Fig.~\ref{sideband} we present the intercycle factor for case A (see table \ref{t1}) as a function of the photoelectron energy and $\cos \theta$ 
at two energy ranges with 10 sidebands each.
In Fig.~\ref{sideband}(a) we show that the sideband energies under the dipole approximation do not depend on the emission angle. Contrarily, the inclusion of non-dipole effects produces the inclination of the sidebands [see Fig.~\ref{sideband}(b) and Fig.~\ref{sideband}(c)]. The vertical dashed lines indicate some reference values corresponding to SB peaks for certain orders $n$ within the dipole approximation. 
When the emission is opposite to the IR propagation ($\cos \theta <0$) the SBs are shifted towards higher energies. On the contrary, when $\cos \theta >0$, the shift is in the direction of lower energies.
This results in an emission asymmetry depending on whether the emission direction is parallel or anti-parallel to the propagation direction of the laser. For each $n$ order SB, the forward (-) and backward (+) energy shift is 
\begin{equation}
\Delta E \simeq \mp \frac{ U_p}{c} \sqrt{2 E_n^D},
\label{width}
\end{equation}
respectively. We also observe that, according to Eq. (\ref{width}), as the energy increases, the slope of the sidebands also increases: the sidebands of Fig.~\ref{sideband}(c) are more slanted than in Fig.~\ref{sideband}(b). Thus, the forward-backward asymmetry emission is more noticeable for higher energies. As sidebands are very robust structures throughout the focal volume \cite{Hummert20}, measuring the relative energy shift of sidebands, i.e., 
$\Delta E / \omega = \mp 2 \beta_0  \sqrt{2 E_n^D} $, might be a helpful tool to determine the experimentally elusive intensity of strong and/or low-frequency NIR lasers through the parameter $\beta_0=q  c/2\omega_L $. 

\subsection{Intracycle factor}

In previous works, we have shown that the sideband structures stemming from the intercycle interferences  are modulated by the intracycle factor   \cite{Gramajo18}. 
Thus, in this section, we analyze how the 
non-dipole description affects the $|I(T)|^2$ modulation. 

\begin{figure}
\centering
\includegraphics[angle = 0, width=0.45 \textwidth]{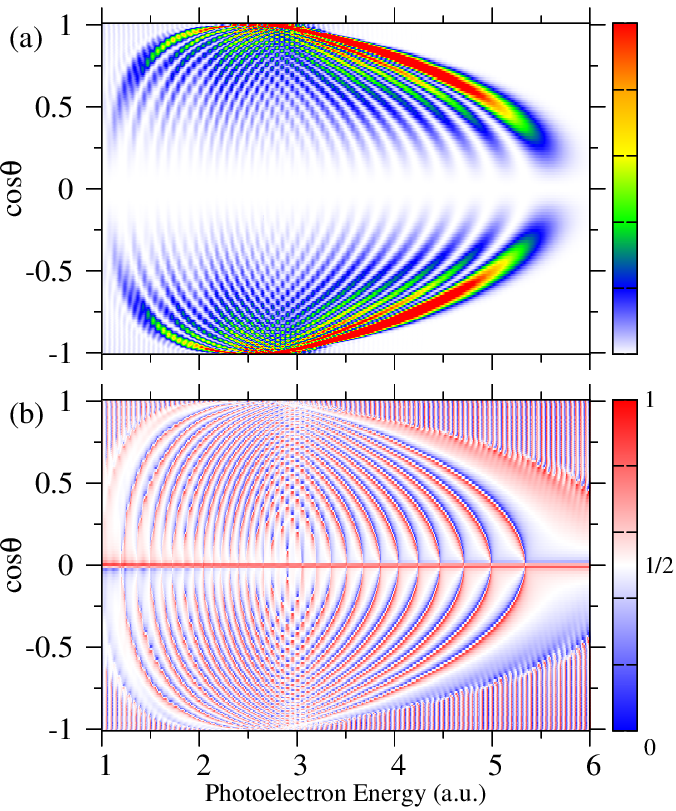}
\caption{(Color online) (a) Non dipole LAPE PES (with $N=1$) of  He($1s$) for case A (see Table~\ref{t1}). 
(b) asymmetry factor \eref{Asy}. 
}
\label{fig_atA}
\end{figure}

\begin{figure}
\centering
\includegraphics[angle = 0, width=0.45 \textwidth]{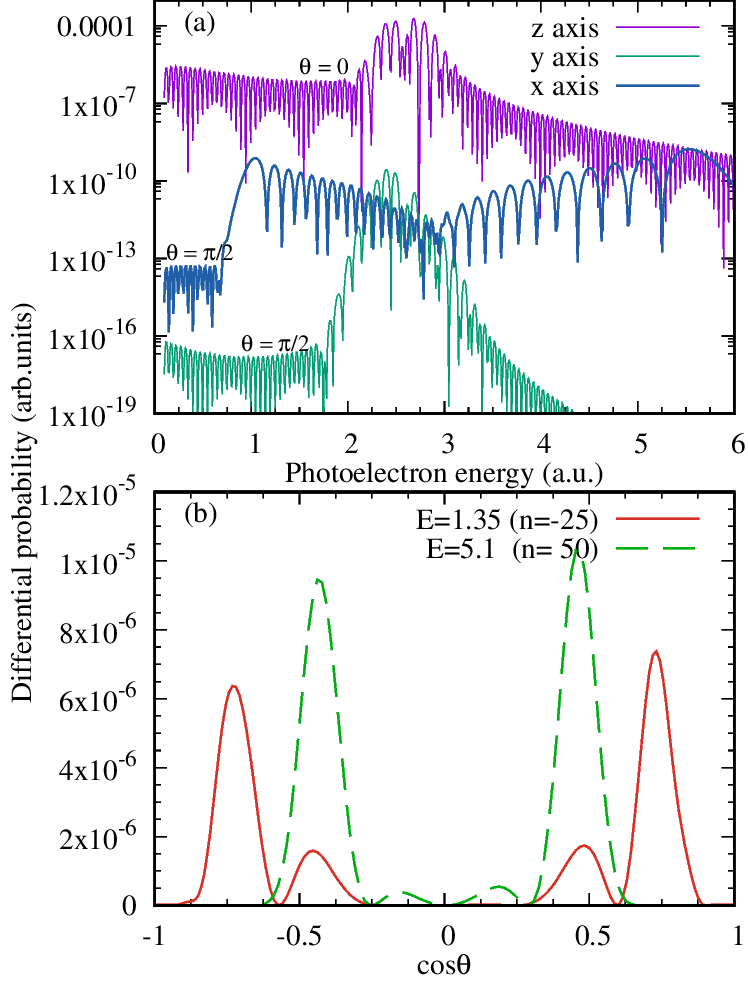}
\caption{(Color online) Intracycle factor for LAPE of He($1s$), non dipole case A (see Table~\ref{t1}). 
(a) at fixed emission angles and (b) at fixed energy values.
}
\label{fig_atA_cuts}
\end{figure}

In Fig. \ref{fig_atA}(a) we show the intracycle factor multiplied by $k$, for the case A (see table \ref{t1}) and for $\vbk = k_z \hat{z} + k_{\perp} \hat{x}$, i.e., emission in the plane $(xz)$. This is equal to the PES  \eref{prob1}, whether $N=1$.  
In Fig.~\ref{fig_atA_cuts} we show the PES for different cuts of the plots of Fig.~\ref{fig_atA}: in Fig.~\ref{fig_atA_cuts}(a) with $\theta$ fixed, and in Fig.~\ref{fig_atA_cuts}(b) at fixed energies as a function of the emission angle. 
We observe that the intracycle factor has a region delimited by certain energy and angle values and it vanishes outside this region. In a previous work \cite{Gramajo18} and within the semiclassical model, this fact was interpreted as a classically allowed region. 
As in the dipole description, inside the allowed region the intracycle factor has fringes. Comparing both, the dipole and non-dipole results, we do not observe  significant qualitative differences in these fringes (not shown). 
However, due to the incorporation of non-dipole terms of order $1/c$, we can expect that the most noticeable variations will be in the areas close to zero emission [white areas in Fig.~\ref{fig_atA}(a)]. 
For that reason, we introduce in Fig.~\ref{fig_atA}(b) the parameter $A$, which quantifies the relative importance of non-dipole effects:
\begin{equation}
A(E,\theta) = \frac{\frac{\mathrm{d}P^{\mathrm{ND}}}{\mathrm{d}Ed\cos \theta }}{\frac{%
\mathrm{d}P^{\mathrm{ND}}}{\mathrm{d}E\mathrm{d}\cos \theta }+\frac{\mathrm{d%
}P^{\mathrm{D}}}{\mathrm{d}E\mathrm{d}\cos \theta }} , \label{Asy}
\end{equation}
where $\rm ND$ and $\rm D$ correspond to the non-dipole and dipole distributions, respectively. The parameter $A(E,\theta)$ will be zero when non-dipole effects are negligible, close to $1$ when they are dominant, and $1/2$ when there are no differences with the dipole case.

Close to the zero-emission areas (white region in Fig.~\ref{fig_atA}(a)), we observe red or blue areas in Fig.~\ref{fig_atA}(b), featuring the non-dipole contributions. 
We also recognize that the blue structures in the lower half plane ($\cos \theta <0$) become red in the upper one ($\cos \theta >0$) and vice versa; which demonstrates the existence of an asymmetry in the forward-backward emission with respect to the IR propagation direction. 
This asymmetry can also be observed, for example, in Fig.~\ref{fig_atA_cuts}(b), where the peaks on the right are higher than their respective ones on the left.

Therefore, we can conclude that 
(i) the forward-backward asymmetry emission does not circumscribe to the inclination of the sidebands. A proper contribution of the intracycle factor is also present;
and (ii) the PES present well-defined regions, where the 
emission probability is considerably higher, similar to the classically allowed regions for dipole-LAPE.
On the borders of these regions, where the PES is close to zero, the $O(1/c)$ contribution will be appreciate. In these sense, it could be motivating to investigate if there are corrections in the classical limits due to non-dipole effects. This is addressed in the next section.

\subsection{Non-dipole classical limits}

\begin{figure}
\centering
\includegraphics[angle = 0, width=0.45 \textwidth]{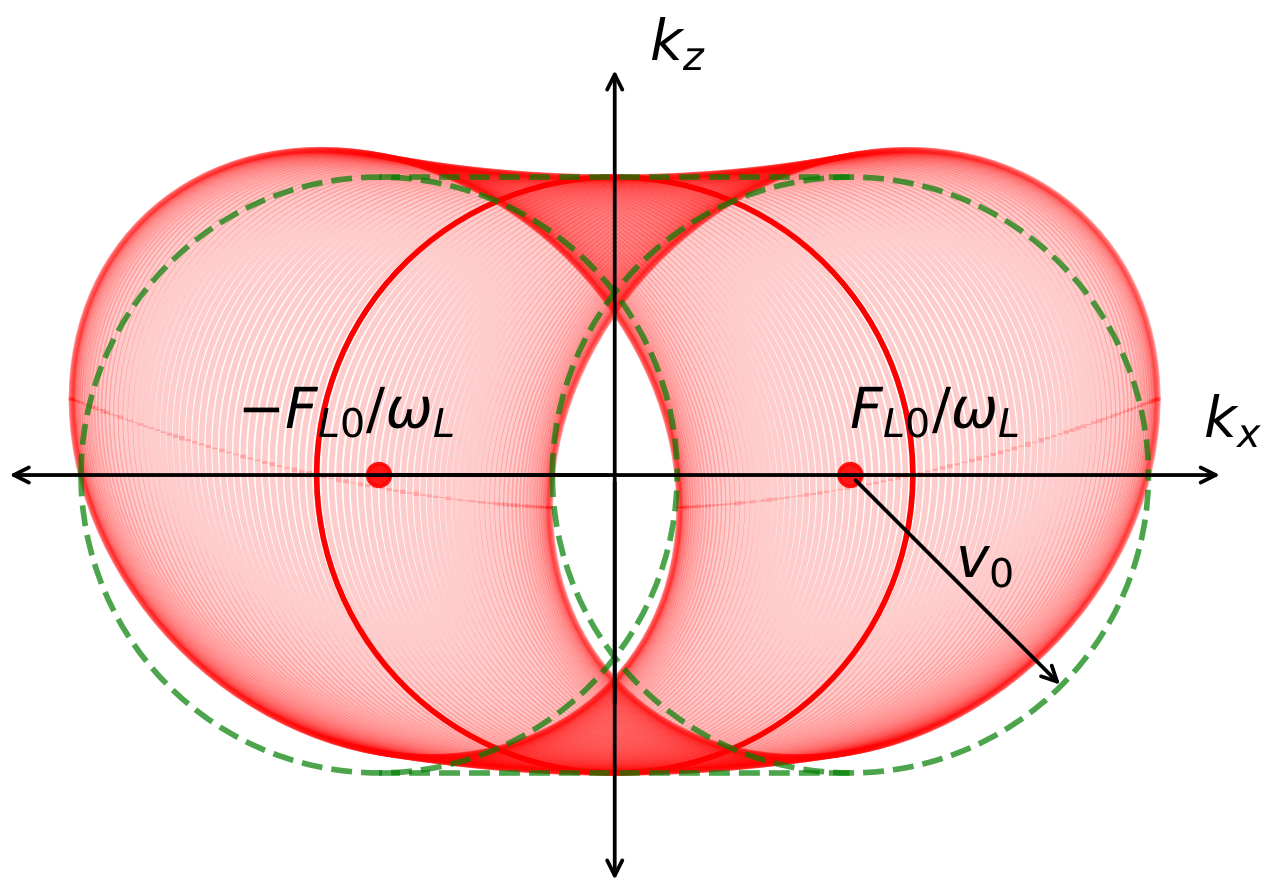}
\caption{(Color online) Schematic classical limits in the momentum space $(k_x,k_z)$. In dashed dark green lines we depict the dipole results (see Ref.~\cite{Gramajo18}). }
\label{fig_limiteclasico}
\end{figure}

The semiclassical model (SCM) consists in solving the time integral of \eref{Tifg} by means of the saddle-point approximation (SPA), where the main contribution occurs for those times $t_s$ for which the action is stationary, i.e., $dS/dt =0$. 
In the non-dipole approach, it means that [see \eref{S1}]:
\begin{equation}
 \boldsymbol{\Pi}^2(\vbk,t_s) = v_0^2,  \label{dsdt0}
\end{equation}
where $v_0^2/2=\omega_X-I_p $ corresponds to the mean energy of the photoelectrons ionized by the XUV pulse in absence of the NIR laser.
Then the transition probability can be written
as a coherent superposition of the amplitudes of all classical
electron trajectories starting from stationary points $t_s$ of the generalized action $S(t)$ with final momentum $\vbk$.
Complex ionization times give rise to non-classical trajectories with exponentially decaying factors and, thus, with minor relevance compared to classical trajectories with real values. 
In other words, those values of the momentum $\mathbf{k}$ satisfying Eq. (\ref{dsdt0}) for real values of $t_s$ define a region of classically allowed momenta.
Considering the $ k_{x},k_{z}$ plane and neglecting terms of the order of $1/c$ within the dipole approximation we get an oscillating circumference, as it has been described in \cite{Gramajo18}.
 The oscillation is harmonic in the direction of the NIR field, i.e., $k_{x}$, with amplitude $F_{L0}/\omega_{L}$ and frequency $\omega_{L}$. In Fig.~\ref{fig_limiteclasico} we show these regions. At $t_s=0$ we get the red circle, while the extreme regions correspond to the two dashed green circumferences. In the present case,  the $O(1/c)$ contribution introduces a small correction. In order to make it visible, we have taken an artificial value of $c=13.7$ to generate the red ellipses in Fig.~(\ref{fig_limiteclasico}). They come from the continuous movement and deformation of the circle as a function of $t_s$ generated by the non-dipole contributions.
 
\begin{figure}
\centering
\includegraphics[angle = 0, width=0.45 \textwidth]{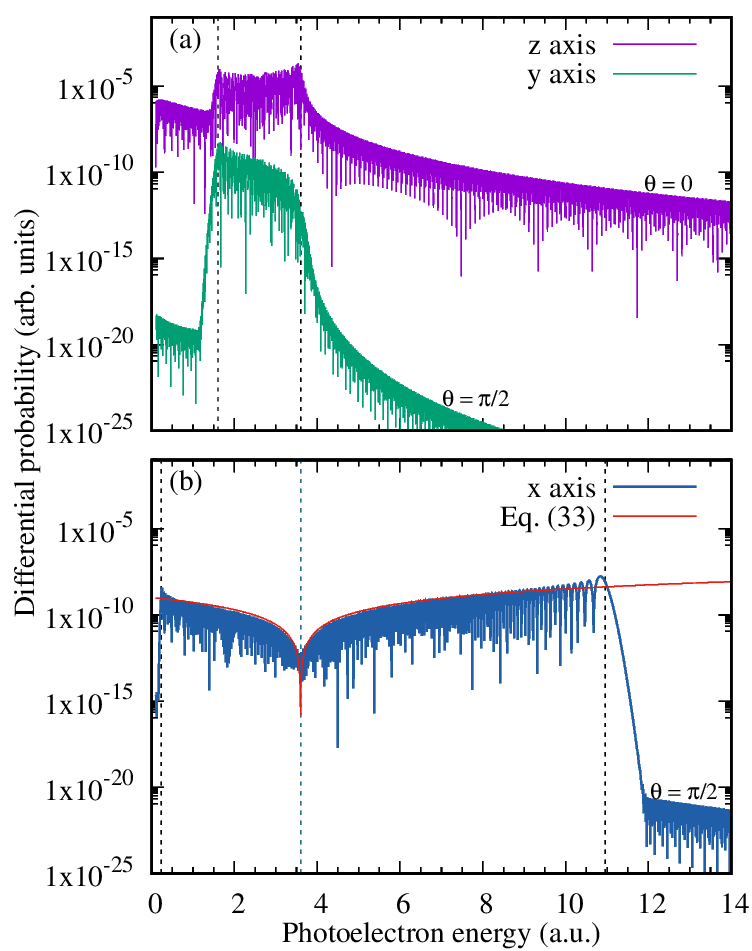}

\caption{(Color online) Intracycle factor for case D as a function of the photoelectron energy at fixed emission angles. In (a) the photoelectron momentum is $\vbk= k\hat{z}$ (in violet line)  and $\vbk= k\hat{y}$ (in green line and several orders of magnitude lower). The black dashed vertical lines indicate the semiclassical limit at $E_\textrm{low}=1.6 $  and $E_\textrm{up}= 3.6$ a.u.  
In (b) $\vbk= k\hat{x}$ in a blue solid line and the quadratic approximation \eref{cuadratica} in a red thin line. 
The black dashed vertical lines indicate the semiclassical limit at $E_\textrm{low}=0.2$ 
and $E_\textrm{up}=11$ a.u. 
at $v_0^2/2=3.6$ (blue dashed vertical line) the PES shows a minimum. 
}
\label{fig_atD}
\end{figure}

As it was shown in \cite{Gramajo18}, and comparing the quantum SFA and TDSE results, the SCM gives an excellent prediction where the PES is non-negligible. We can expect, then, a small non-dipole variation in the classical limits.  In order to analyze it, we consider 
the three principal directions: $\hat{x},\hat{y}$ and $\hat{z}$. 
In Fig.~\ref{fig_atD} we show the intracycle factor for the case D (see table \ref{t1}) for emission in the $\hat{z}$ and $\hat{y}$ direction in Fig.~\ref{fig_atD}(a), and $\hat{x}$ in Fig.~\ref{fig_atD}(b). The vertical dashed lines indicate the classical limits obtained as follows. The  \eref{dsdt0} (neglecting $1/c^2$ terms in \eref{Pi2}) gives 
\begin{eqnarray}
v_0^2 &=& k^2 + A_L^2 + \frac{A_L^2 k}{c}  \qquad  \textrm{ if (i)} \quad \vbk= k\hat{z}  \label{lim_z} \\
v_0^2 &=&   k^2 + A_L^2 \qquad \qquad \quad
\textrm{ if (ii)}\quad  \vbk= k\hat{y} \\
v_0^2 &=&  (k + A_L)^2 \qquad \qquad 
\textrm{ if (iii)}\quad  \vbk= k\hat{x}. \label{pivox}
  \end{eqnarray}
The maxima and minima classically allowed $k$-values are those for
which the above equations have extreme values (maxima and minima) of the field $A_L(\omega_L t)$ or $A_L^2(\omega_L t)$. So, in the second (ii) case, the values $0$ and $(F_{L0}/\omega_L)^2 = 4 U_p$ minimizes and maximizes the field $A_L^2(\omega_L t)$, respectively, giving rise to the classical limits: 
$E_\textrm{low} \simeq (v_0^2 - 4 U_p)/2$  and $E_\textrm{up} = v_0^2/2$. 
For case (iii), the limits are $E_\textrm{low,up} = (v_0 \mp F_{L0}/\omega_L)^2/2$.
These two cases (ii) and (iii) coincide exactly with those expected for the dipole approximation (see \cite{Gramajo18} for emission ``perpendicular'' and ``parallel'' to the dressing NIR field).
Instead, for case (i) there is a small non-dipole correction to the dipole classical limit at low
energy. Whereas $E_{\textrm{up}} = v_0^2/2$, the lower limit is 
\begin{equation}
    E_{\textrm{low}} =\frac{ v_0^2}{2} -2 U_p (1 \pm \frac{1}{c}\sqrt{v_0^2-4 U_p}). 
    \label{Elowi}
\end{equation}
We observe in Fig.~\ref{fig_atD} that, effectively, the larger emission probability is restricted to the SCM range, delimited by the classical borders $E_\textrm{low,up}$.

The most striking difference compared with the dipole results lies in the fact that the emission in $\hat{x}$ is not forbidden, see Fig.~\ref{fig_atD}(b). 
Indeed, in the dipole approximation, the dipole element $\vbd(\vbk+\vbA)$ is orthogonal to the XUV polarization vector in $\hat{z}$, then emission in the $\hat{x}$ direction is forbidden. Instead, according to \eref{Apen_ell}, beyond the dipole approximation, we can expect some contribution of the order $O($1/c$)$. Hence, all the contribution in this direction is purely non-dipole. 
Furthermore, we can observe a very noticeable structure: a minimum  at $E=v_0^2/2$. This structure can be easily understood in light of the semiclassical model: since
the principal contribution to the temporal integral occurs at real times $t_s$ verifying \eref{dsdt0}, we can approximate $\ell(t)$ by:
\begin{eqnarray}
\ell(t_s) & \propto &  \frac{1}{c(v_0^2 + \alpha^2)^3} \Big(k A_L(\omega_L t_s) + \frac{A_L^2(\omega_L t_s)}{2} \Big) \nonumber \\
 & \propto & \frac{v_0^2}{2} - \frac{k^2}{2},
\end{eqnarray}
where we have combined Eqs.~(\ref{dsdt0}), (\ref{pivox}) and (\ref{Apen_ell}).
We note that this expression does not depend on the time $t_s$ but on the energy $E=k^2/2$ and vanishes at the particular value $E= v_0^2/2$. Since it represents a zero of the matrix element,  we can call it a `Cooper-like minimum'.
Then the intracycle factor is proportional to a quadratic function of the energy that vanishes at the Cooper-like minimum:
\begin{equation}
|I(T)|^2 \propto  | \ell(t_s)|^2 \sim (\frac{v_0^2}{2} - E)^2 \label{cuadratica}
\end{equation}
This quadratic function is plotted as a red thin line in Fig.~\ref{fig_atD}(b), with an arbitrary normalization constant.  The agreement is striking.

\begin{figure}
\centering
\includegraphics[angle = 0, width=0.45 \textwidth]{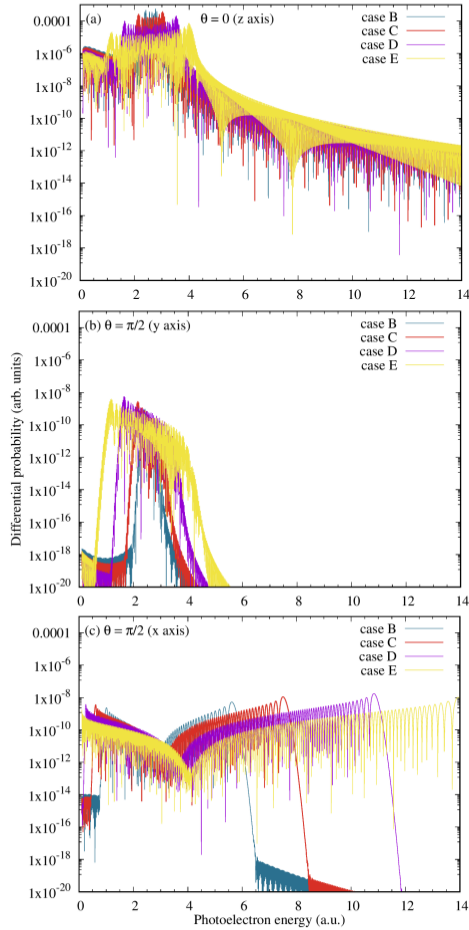}

\caption{(Color online) Intracycle factor at the emission direction (a) $\vbk= k \hat{z}$, (b)  $\vbk= k \hat{y}$ and  (c) 
 $\vbk= k \hat{x}$, for different laser parameters according to cases B, C, D and E (see Table \ref{t1}).
}
\label{fig_vsbeta0}
\end{figure}

Now, we want to explore the behavior of the intracycle factor and its relationship with the classical limits as a function of the laser configuration by increasing the $q$ parameter,
which indicates a growth of non-dipole contributions. 
For that purpose, we consider cases B, C, D, and E of Table~\ref{t1}.
In Fig.~\ref{fig_vsbeta0} we present the spectra for the three principal emission directions as previously considered.
We observe that the spectra remain limited by the classical values and they widen as $q$ increases.
We also see that the qualitative shape of the spectra does not change and, notoriously, the Cooper-like minimum also persists (see Fig.~\ref{fig_vsbeta0}(c)) at different energy positions as the XUV frequency varies.

\begin{figure}
\centering
\includegraphics[angle = 0, width=0.45 \textwidth]{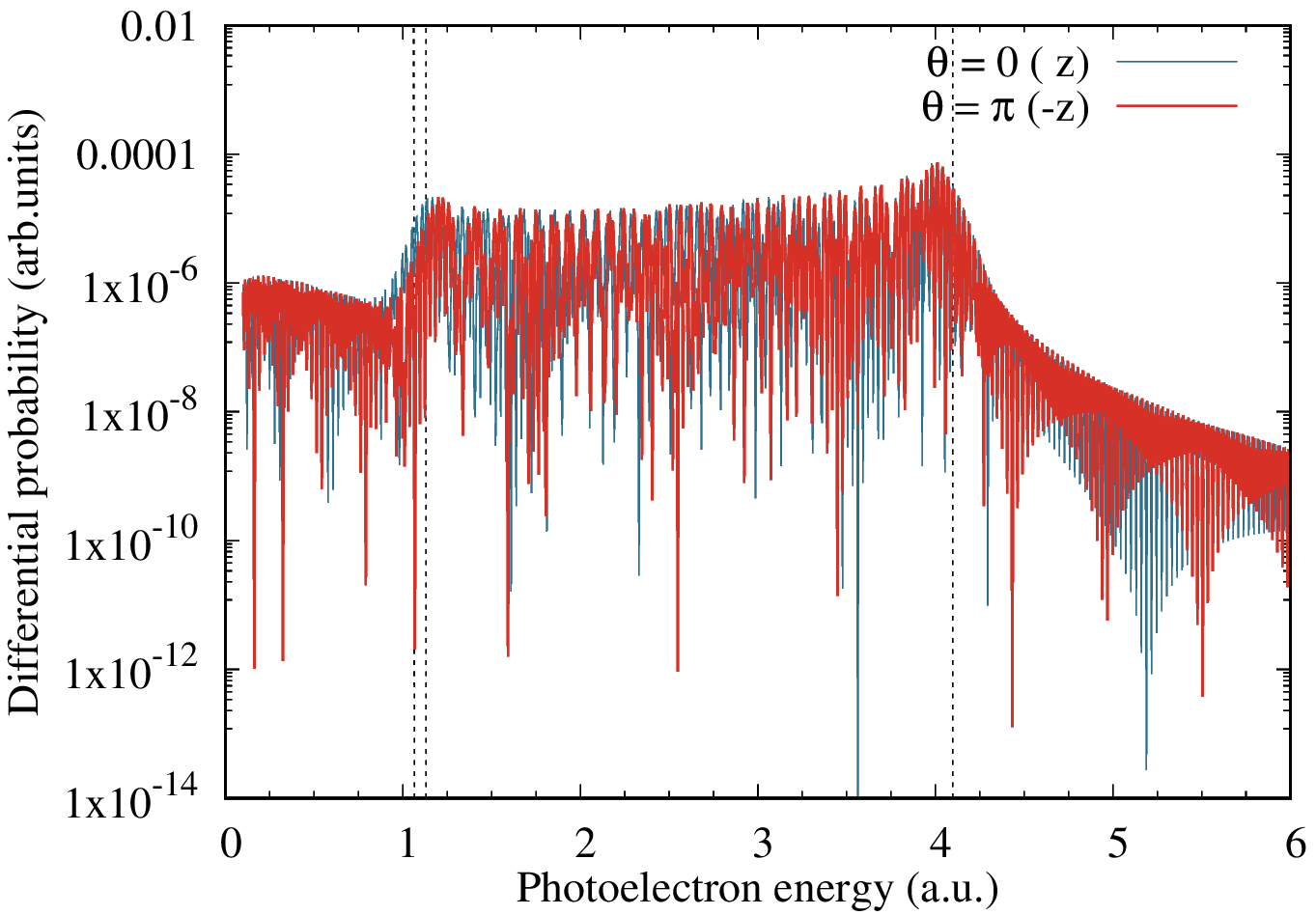}
\caption{(Color online) Intracycle factor at the emission direction $\hat{z}$ (blue) and $-\hat{z}$ (red) for case E (see Table~\ref{t1}). The dashed vertical lines indicate the classical limits at $E_\textrm{low+}=1.06$, $E_\textrm{low-}=1.13$ and $E_\textrm{up}=4.09$ a.u.
}
\label{fig_atE_asymmetry}
\end{figure}

Finally, we want to study if there is some kind of forward-backward asymmetry  in the classical limits.  For that, we compare  the emission in the parallel ($\hat{z}$)  and antiparallel ($-\hat{z}$) directions with respect to the IR propagation direction.
In Fig.~\ref{fig_atE_asymmetry} we compare both situations for case E.
A very good qualitative agreement is observed at the entire range, except close to the lower classical limit.
 According to \eref{lim_z}\footnote{the third term of this equation is negative for antiparallel emission.}, the lower classical limits depend on the emission direction (forward of backward, i.e., $\pm \hat{z}$) according to Eq. (\ref{Elowi}).
 This small difference can be observed comparing the forward and backward spectra: the blue curve ``rises'' before (at $E_\textrm{low$+$}=1.06$) than the red one (at $E_\textrm{low$-$}=1.13$).
Within the dipole approximation, both curves coincide exactly at the lower classical limit $E_\textrm{low}=1.09$, which lies precisely in the middle of both (not shown).

Summarizing this section, we can say that the classical limits accurately determine the energy and emission angle range. These limits are slightly affected by non-dipole effects. They appear as a forward-backward asymmetric emission. Moreover, beyond the dipole approximation, there is no restriction in the emission in the $\hat{x}$ direction, and in this case the semiclassical model predicts perfectly the shape of the spectra and the presence of a Cooper-like minimum at energy equal to $v_0^2/2$. 

\section{Conclusions}
\label{conclusions}

 We have studied the laser-assisted photoemission (LAPE) process in a nonrelativistic SFA description including non-dipole corrections.
 Due to the periodicity properties, and in the same spirit of our previous work \cite{DellaPicca2020}, we can rewrite the PES as a function of two principal contributions:  the intra- and the intercycle factors.
 We have analyzed each factor and the modifications introduced in each of them by the nondipole effects.
In order to consider a concrete study case, 
we have analyzed the LAPE of He($1s$) in a particular geometrical arrangement of the XUV and IR polarization vectors and IR propagation direction, and for several IR laser parameters.
As a result of the intercycle interference, the sideband pattern exhibits an angle dependence (inclination), which increases with energy.
The classical allowed angle-energy region, previously investigated within the dipole approximation is sensitive to corrections of order $O(c^{-1})$.
As a particular point, we have found that the emission restriction at $\hat{x}$  direction is broken in the non-dipole approach. The non-dipole semiclassical model shows  an excellent agreement with the SFA results and gives a useful interpretation of its most notorious structure, the Cooper-like minimum. 
In both intra- and intercycle factors and also in the borders of the classically allowed region, we have observed and characterized a forward-backward asymmetry originated in the existence of a privileged direction that is the IR propagation one.  A more detailed analysis of this asymmetry deserves to be treated in future works.

\appendix
\section{Appendix: Transition matrix dipole element} \label{apen_dipol}

The dipole transition element is defined as:
\begin{equation}
\vbd(\vbv) = \frac{1}{(2\pi)^{3/2}} \int \rmd \vbr \, \exp(-i \vbv\cdot \vbr)\,  \vbr \, \phi_i(\vbr),
\end{equation}
where $\phi_i$ is a hydrogen-like bound state.  For the case of a hydrogenic $1s$ state, we can write
\begin{eqnarray}
\vbd (\vbv) &=& -\frac{i}{\pi}\, 2^{7/2}\alpha^{5/2}\, 
\frac{ \vbv}{ (v^2+\alpha^2)^3}  \label{1s}
\end{eqnarray}
where $\alpha = \sqrt{2I_p}$.
In the present work we have considered the ionization energy  $I_p =24.587$ eV (= 0.90356 a.u.)  for the $1s$ state of He ($Z_\textrm{eff}=1.34429$).

In order to compute the presented results it was necessary to evaluate $\ell (t)$ (\eref{lt}) considering  $\hat{\varepsilon}_X=\hat{z}$ and $\hat{\varepsilon}_L=\hat{x}$. Thus, 
\begin{eqnarray}
\ell (t) &=& -\frac{F_{X0}}{\pi} (2 \alpha)^{5/2} \frac{ \hat{z} \cdot  \boldsymbol{\Pi}(\vbk,t)}{( \boldsymbol{\Pi}^2(\vbk,t)  + \alpha^2)^3 } \label{Apen_ell}  \\
&=&\frac{ - F_{X0} (2 \alpha)^{5/2}}{ ( \boldsymbol{\Pi}^2(\vbk,t)  + \alpha^2)^3 \pi } [ \vbk\cdot\hat{z}  + \frac{\vbk\cdot\vbA_L(\omega_L t)}{c} + \frac{\vbA_L^2(\omega_L t)}{2c} ] \nonumber.
\end{eqnarray}
where $\boldsymbol{\Pi^2}$ is detailed in \eref{Pi2}.

\begin{acknowledgments}
This work is supported by PICT 2020-01755, 2020-01434, and PICT-2017-2945 of ANPCyT (Argentina), PIP 2022-2024 11220210100468CO of Conicet (Argentina).
M.F.C. acknowledges financial support from the Guangdong Province Science and Technology Major Project (Future functional materials under extreme conditions - 2021B0301030005).
\end{acknowledgments}

\bibliography{biblio}

\end{document}